\documentclass{article}
\usepackage{ijcai23}

\usepackage{times}
\usepackage{soul}
\usepackage{url}
\usepackage[hidelinks]{hyperref}
\usepackage[utf8]{inputenc}
\usepackage[small]{caption}
\usepackage{graphicx}
\usepackage{amsthm}
\usepackage{stfloats}
\usepackage{tabularx}
\usepackage{amsmath}

\usepackage{subcaption}
\usepackage{cleveref}
\usepackage{booktabs}
\usepackage{algorithm}
\usepackage{algorithmic}
\usepackage[switch]{lineno}

\usepackage{amssymb}
\usepackage{accents}

\title{SCALA: Sparsification-based Contrastive Learning for Anomaly Detection on Attributed Networks}

 \author{
 Enbo He$^{1,*}$
 \and
 Yitong Hao$^{1,*}$\and
 Yue Zhang$^{1,\dag}$\and
 Guisheng Yin$^{1,\dag}$\And
 Lina Yao$^2$
 \affiliations
 $^{1}$College of Computer Science and Technology, Harbin Engineering University\\
 $^{2}$School of Computer Science and Engineering, University of New South Wales\\
 \emails
\{enochhe, haoyitong\}@hrbeu.edu.cn, 
zycg87@sina.com,
yinguishengabc@163.com,
lina.yao@unsw.edu.au
 }

\begin{document}
\maketitle
\begin{abstract}
    Anomaly detection on attributed networks aims to find the nodes whose behaviors are significantly different from other majority nodes. Generally, network data contains information about relationships between entities, and the condition of a relationship can provide important hints for finding node anomalies. Therefore, how to comprehensively model complex interaction patterns in networks is still a major focus. It can be observed that anomalies in networks violate the homophily assumption. However, most existing studies only consider this phenomenon obliquely rather than explicitly. Besides, the node representation of normal entities can be perturbed easily by the noise relationships introduced by anomalous nodes. To address the above issues, we present a novel contrastive learning framework for anomaly detection on attributed networks, \textbf{SCALA}, aiming to improve the embedding quality of the network and provide a new measurement of qualifying the anomaly score for each node by introducing sparsification into the conventional contrastive method. Extensive experiments are conducted on five real-world datasets and the results show that SCALA consistently outperforms all baseline methods significantly.
\end{abstract}

\section{Introduction}

Tasks on networks have drawn much attention in recent years \cite{network}, particularly for anomaly detection on the network, because there are many complex interaction patterns between entities hard to process \cite{anomaly}. Unlike in other domains, the anomaly detection in networks requires the identification of multiple anomaly types, including attribute anomaly and structure anomaly \cite{dominant}. The former describes the node with unusual condition of attributes and the latter means the node that has the abnormal situation of the network topology. How to simultaneously identify these anomalies remains a major current focus.

There have been many studies on network anomaly detection. In the early research, many shallow methods were proposed. Their main idea is largely about designing metrics that evaluate the degree of abnormality. For instance, \cite{lof} design a metric that describes the isolation level of nodes. \cite{scan} take the difficulty of clustering a node into consideration. \cite{radar} and \cite{anomalous} deploy the residual analysis to measure the anomaly score for each node. These methods heavily rely on human experience in anomaly judgment and have limited ability to utilize the complex relationships of the network. Hence their actual capacity in anomaly detection is limited.

Thanks to the increasing development of GNN which provides a tool to extract the topology and attribute knowledge at the same time \cite{gnn}. Many researchers introduced GNN into anomaly detection on attributed networks \cite{anomaly}. In general, there are three main categories of GNN methods. First, the Auto-Encoder based method assumes that the attribute and the topology pattern of a normal node can be correctly reconstructed by the decoder while the abnormal nodes cannot \cite{dominant}. Second, the One-Class SVM based method believes that all the normal nodes can be mapped in the boundary of a hypersphere, while the abnormal node will be expelled out of the boundary \cite{ocgnn}. At last, the Contrastive Learning based (CL-based) method determines the degree of node anomaly by calculating the agreement score between the target node and its surrounding neighbor nodes. For normal nodes, their dense embeddings should maintain a matching condition with their positive subgraph embeddings, while abnormal nodes do not retain such a condition \cite{cola}. A number of studies have shown that the CL-based method is currently the best method of all, because it can comprehensively and intuitively leverage the matching relationship between the target node and its neighbors \cite{cola,anemone,slgad,subcr,gradate}.

\begin{figure}[t]
	\centering
	\setlength{\abovecaptionskip}{-0.01cm}
	\setlength{\belowcaptionskip}{-0.3cm}
        \vspace{-0.5cm}
	\begin{subfigure}{0.48\linewidth}
		\centering
		\setlength{\abovecaptionskip}{-0.01cm}
		\includegraphics[width=1\linewidth]{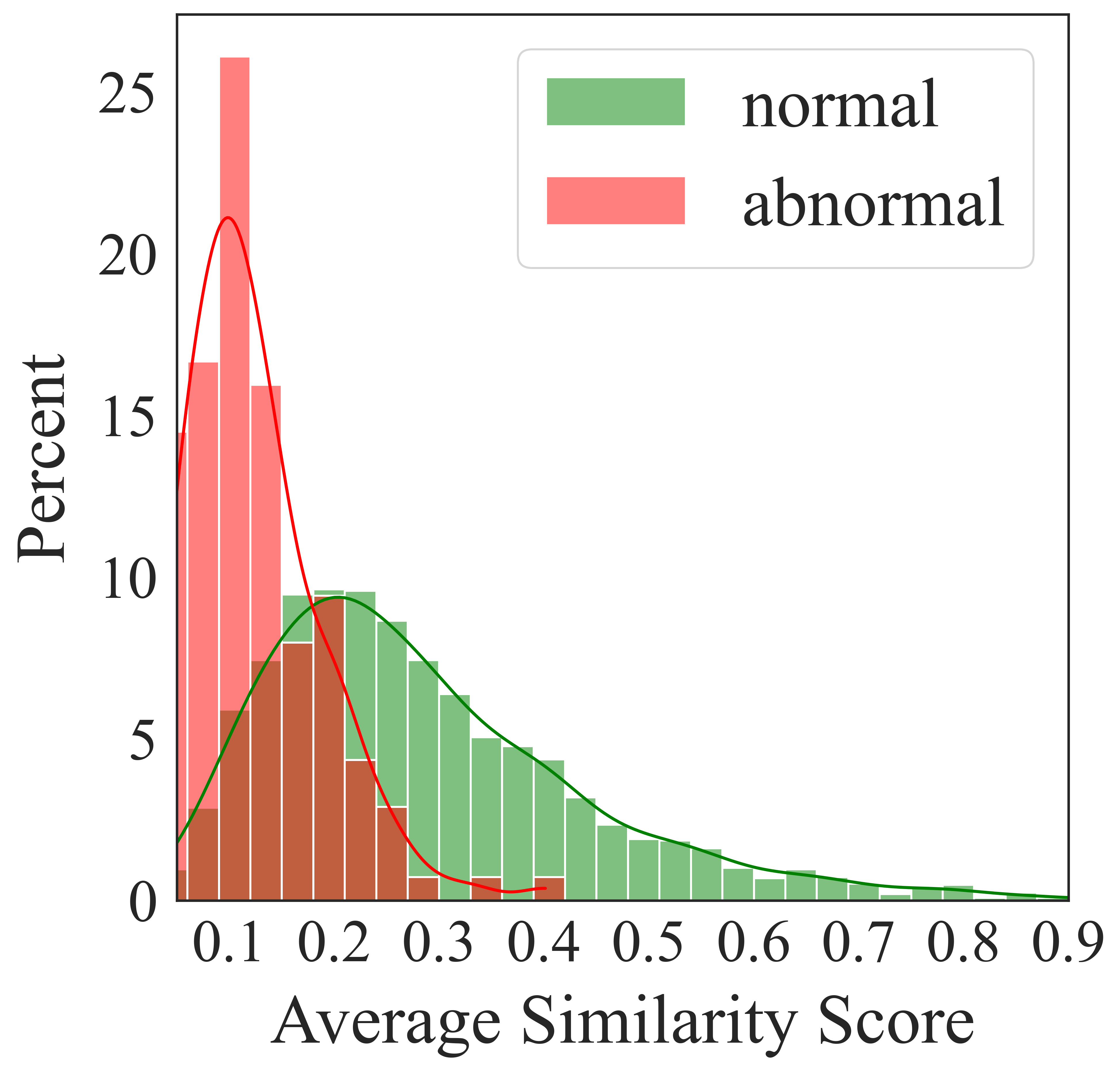}
		\caption{}
	\end{subfigure}
	\centering
	\begin{subfigure}{0.48\linewidth}
		\centering
		\setlength{\abovecaptionskip}{-0.01cm}
		\includegraphics[width=1\linewidth,trim=50 8 0 0,clip]{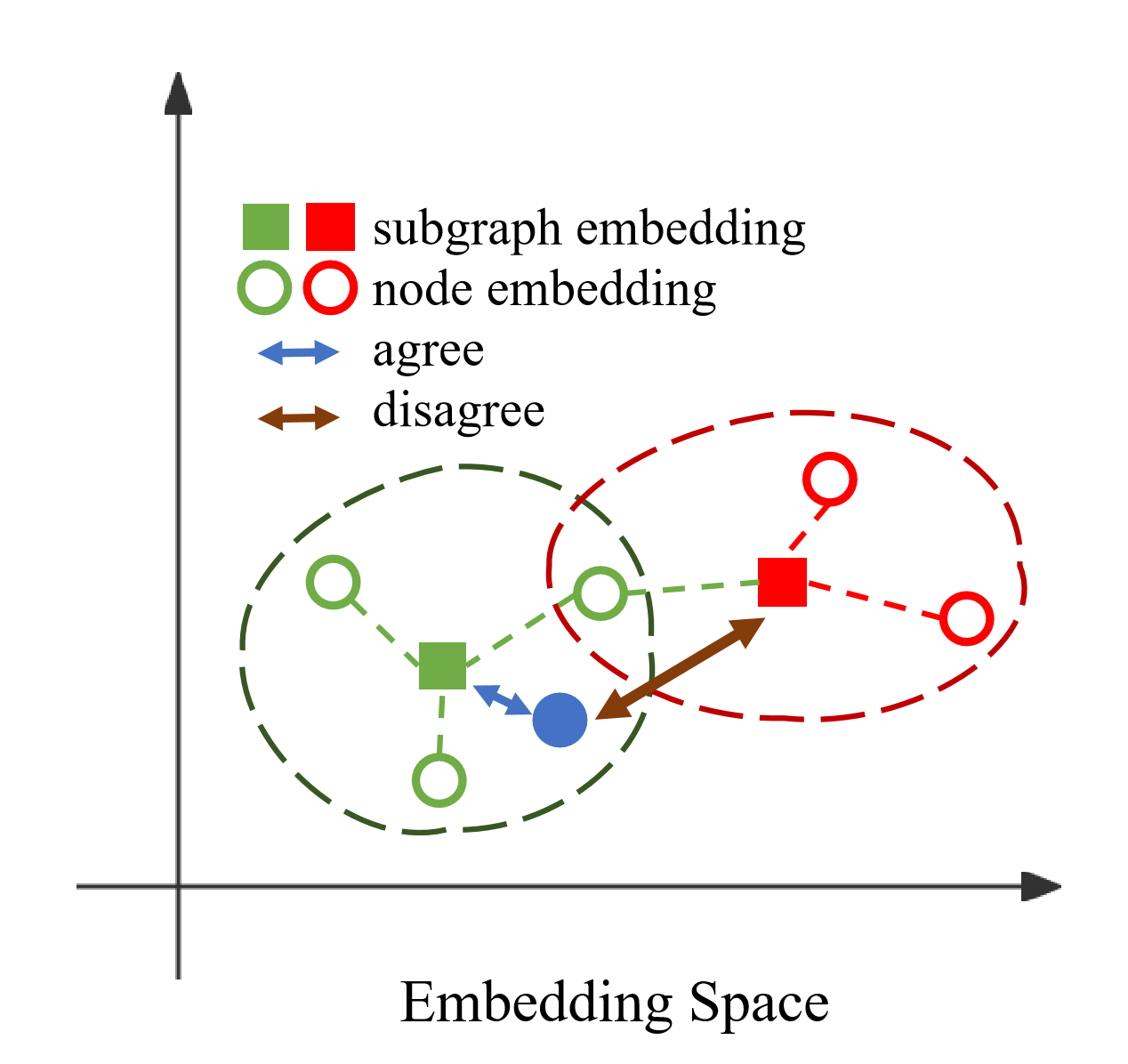}
		\caption{}
	\end{subfigure}
    \caption{Illustration of the property and the impact of anomalies. (a) The average similarity score of normal and abnormal nodes are calculated by Eq.(1) respectively. We display their percentage distribution by using a histogram. It can be seen that abnormal nodes are more concentrated in areas with low homogeneity. (b) The green and red circle represent the correct and wrong subgraph sampling respectively corresponding to the target node which is presented by the blue dot. We can see that the abnormal node in the subgraph will lead disagreement between the target node and the subgraph embedding.}
    \label{fig: Figure 1}
\end{figure}

It is generally accepted that most nodes of the real-world network satisfy the homophily assumption, which means that connected nodes tend to have a high degree of similarity \cite{homophily}. However, as shown in Figure \ref{fig: Figure 1}(a), the abnormal nodes don't. Surprisingly, using this discipline as the criteria for spotting anomalies has remained underexplored in previous studies. Besides, as illustrated in Figure \ref{fig: Figure 1}(b), the noise information of abnormal nodes will cause divergence when obtaining the embedding of the subgraph because most of the CL-based methods inevitably pick the abnormal node during sampling the contrastive pair. Therefore, the agreement calculation will have an under-best performance. 
 
The above-mentioned issues can be effectively eased by conducting sparsification on networks which filters the abnormal relationships based on the similarity between the nodes. Because it can provide the prior knowledge of the network interaction patterns, aiding the model in selecting the appropriate edges for message passing. Besides, if a node is connected by more anomalous relationships, the more suspicious for the node to be an anomaly. Motivated by such idea, a new way of qualifying the anomaly of each node is proposed. In this paper, we present an unsupervised multi-view contrastive learning framework, titled \textbf{SCALA} (\textbf{S}parsification-based \textbf{C}ontr\textbf{A}stive \textbf{L}earning for \textbf{A}nomaly Detection on Attributed Networks). In SCALA, the introduced sparsification method is seen as the view augmentation in the contrastive learning framework. The contrastive anomaly detection is conducted on both views with different readout strategies. The modules on two views are then integrated to calculate the final anomaly score. We summarize the main contributions of this paper as follows:
\begin{itemize}
\item A novel contrastive learning anomaly detection scheme for attributed networks is proposed.  We first utilize sparsification as a method of view enhancement which effectively reduces the divergence on graph-level embedding caused by abnormal nodes.
\item We creatively propose a new way to identify anomalies which complement the structural information neglected by the conventional CL-based method.
\item Extensive experiments are conducted on five real-world datasets, whose results show that \textbf{SCALA} consistently outperforms all baseline methods significantly.
\end{itemize}

\section{Related Work}

\subsection{Anomaly Detection on Attributed Network}

In the early research, shallow methods were proposed to tackle the anomaly detection task. \cite{radar} and \cite{anomalous} adopt matrix decomposition methods and analyze the residual attributes to qualify the anomaly degree. \cite{lof} and \cite{amen} both design novel metrics that measure the abnormality. \cite{scan} uses the difficulty degree of clustering as the score of judge anomalies. Despite their simplicity, these shallow methods are incapable of modeling sophisticated network interaction patterns.

With the rapid development of deep learning, researchers also tried to adopt deep learning techniques to detect anomalies in networks. For instance, \cite{dominant} adopts an autoencoder to spot anomalies by measuring the reconstruction errors of nodes. \cite{guide} explicitly introduces the high-order motif structure information into the autoencoder scheme. \cite{comga} considers the community condition of a node with a tailored GCN. \cite{ocgnn} effectively combines the outliers detection ability of one-class SVM and the representation capacity of GCN. \cite{aagnn} takes full advantage of the deviation of nodes from the background community to induce the normal node embeddings. Moreover, \cite{cola} first adopted the contrastive approach to anomaly detection on the attributed networks. And many schemes based on CoLA are proposed. \cite{anemone} deploy both node-node and node-subgraph contrast pairs in the module. \cite{gradate} investigate the efficiency of subgraph-subgraph contrast pairs. \cite{slgad} and \cite{subcr} simultaneously involve information reconstruction and contrastive learning in the training phase and score inference phase. \cite{arise} combine the substructure anomaly discerning and the contrastive method. \cite{nlgad} take the reliability of the normal node into consideration.

\subsection{Graph Contrastive Learning}

Contrastive learning aims to learn networks that map representation in positive pair instances similar to each other, and that embeddings in positive pair instances disagree with each other. \cite{graphcl,ssl} systematically studies the impact of many transformation methods applied in the contrastive learning, including node attribute masking, random edge perturbation, random node dropping, and random walk-based subgraph sampling. \cite{mvgrl} proves the effectiveness of GDN (Graph diffusion networks) as a view augmentation method. \cite{dgi}, \cite{infograph}, and \cite{mvgrl} validate adequately the Ego-nets sampling as a favorable sample method. By contrast, \cite{gcc} considered that random walk sampling is a stronger method. Most CL-based anomaly detection frameworks employ one or more above-mentioned strategies such that the complicated patterns can be fully modeled.

\section{Problem Formulation}
In this paper, the bold lowercase letter (e.g., $\mathbf{x}$) and uppercase letter (e.g., $\mathbf{X}$) are used to indicate vectors and matrices, respectively. The calligraphic fonts (e.g., $\mathcal{V} $) are used to denote sets. The $i^{th}$ row of a matrix $\mathbf{X}$ is denoted by $\mathbf{x}_i$ and the $\left (i,j \right) ^{th}$ element of $\mathbf{X}$ is denoted by $\mathbf{X}_{i,j}$. 

\textbf{Definition 1.} Attributed Networks: Given a attributed network $\mathcal{G} =\left ( \mathcal{V}, \mathcal{E}, \mathbf{X} \right) $, where $\mathcal{V} =\left\{ v_1, \cdots, v_n \right\}$ is the set of nodes, the number of nodes $|\mathcal{V} |$ is $n$, and the $\mathcal{E} =\left\{ e_1, \cdots , e_m \right\}  $ is the set of edges and the number of edges $|\mathcal{E} |$ is $m$. The topology information is presented by adjacent matrix $\mathbf{A}\in \mathbb{R} ^{n\times n}$, if there is a edge connecting the $i^{th}$ and $j^{th}$ node, $\mathbf{A}_{i,j}=1$. Otherwise, $\mathbf{A}_{i,j}=0$. The attribute matrix is denoted by $\mathbf{X}\in \mathbb{R} ^{n\times f}$ and the attribute of $i^{th}$ node is indicated by $\mathbf{x}_i\in \mathbb{R} ^f$ where $f$ is the dimension of the attribute vector.

\begin{figure*}[t]
    \centering
    \setlength{\belowcaptionskip}{-0.2cm}
    \includegraphics[width=1\textwidth]{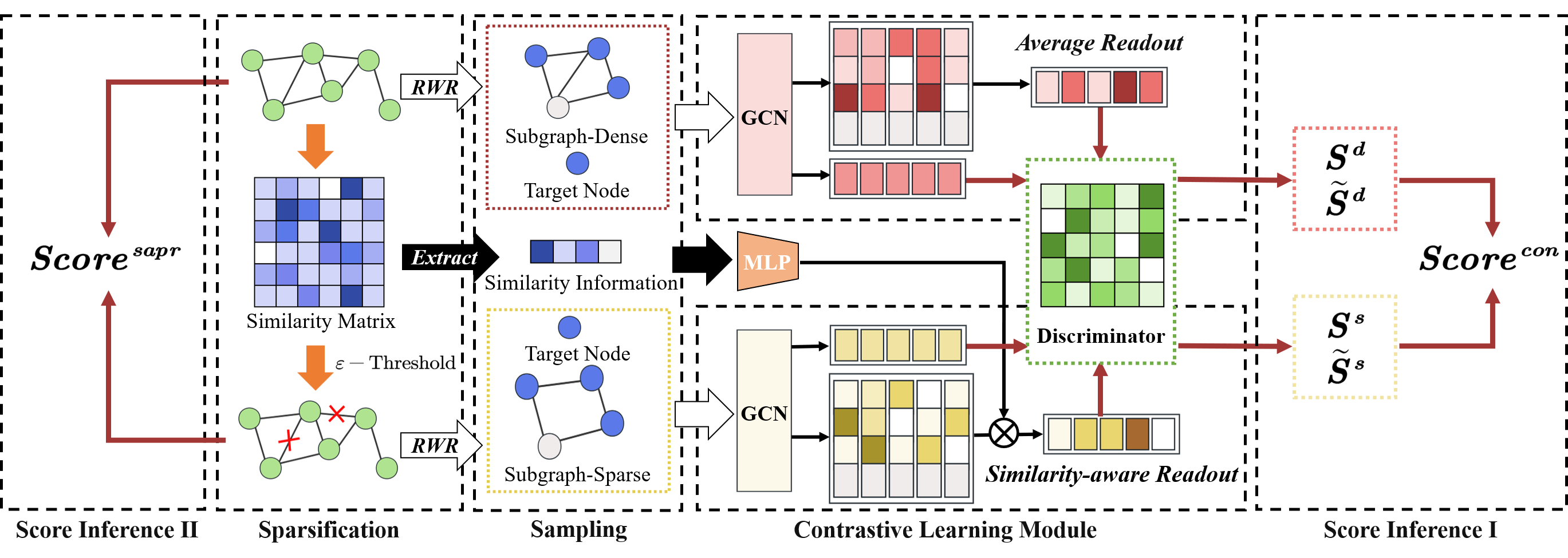}
    \caption{The framework of the SCALA.}
    \label{fig: Figure 2}
\end{figure*}

\textbf{Problem 1.} Anomaly Detection on Attributed Networks: For a attributed network $\mathcal{G} =\left( \mathcal{V}, \mathcal{E}, \mathbf{X} \right)$, our aim is to learn a scoring function $score\left( \cdot \right) $ for qualifying the degree of abnormality. To be specific, the larger the anomaly score $s_i$ indicates the node $v_i$ is more likely to be an anomaly.

\section{Methodology}

In this section, we introduce the proposed method in detail. The overall pipeline and algorithm are illustrated in Figure \ref{fig: Figure 2} and Algorithm \ref{alg:algorithm}. SCALA is mainly consists of four components. To be specific, it first calculates the similarity matrix of the node attribute, which is efficient in capturing the relationship of the whole network and the sparsification is conducted to obtain the \textit{spar-view} (abbreviation for sparse-view), while the origin network is called the \textit{dense-view}. After that, SCALA sample the subgraph around the target node for both view. Then, the target node and its corresponding subgraph are fed into contrastive modules of two views respectively to calculate the discriminative scores. With the attention to ease the negative impact on the quality of graph-level embedding, different readout strategies are operated in two views. After training, SCALA finally integrates scores of the sparsification and the contrastive learning module to gain a final anomaly score.

\subsection{Graph Augmentation via graph sparsification and Subgraph Sampling}

In our approach, the graph sparsification is conducted to obtain a new view due to such a method can improve the robustness of embeddings. In order to acquire the node-subgraph pair for contrastive learning, a random walk based sampling method is adopted, which is proven to be effective \cite{graphcl}.   

\subsubsection{Graph Sparsification}

In order to form the \textit{spar-view}, we adopted the tailored $\varepsilon$-threshold sparsification by calculating the similarity matrix $\mathbf{S}^{\mathcal{G}}$ of the original attribute network \cite{sparsity,sparsity1}, which can be formulated as: 
\begin{equation}
\label{eq1}
	\mathbf{S}_{i,j}^{\mathcal{G}}=sim\left( v_i,v_j \right)
\end{equation}
where $sim\left ( v_i, v_j \right) $ denotes the similarity score of two attributes of nodes. In this paper, dot similarity is adopted which means $sim\left ( v_i, v_j \right) =\mathbf{x}_i\cdot \mathbf{x}_{j}^{T}$. Then a row min-max normalization is performed on the similarity matrix. Furthermore, based on the homophily assumption, if the similarity between two connected nodes is small, such an edge is more suspicious. So we delete the edge whose similarity score is less than a threshold $\varepsilon$ to gain the topology of the \textit{spar-view}, which can be formulated as:
\begin{equation}
\label{eq1}
	\mathbf{A}_{i,j}^{spar}=\left\{ \begin{array}{l}
	1 ,  \quad \mathbf{S}_{i,j}^{\mathcal{G}}>\varepsilon \,\,and\,\,\mathbf{A}_{i,j}=1\\
	0 ,  \quad   otherwise\\
\end{array} \right.
\end{equation}
As mentioned in the introduction section, a node connected by more abnormal edges is more likely to be an anomaly, which provides a new way of qualifying the degree of abnormality. Based on such idea, we propose a new score calculation method by the process of sparsification, which can be formulated as:
\begin{equation}
\label{eq1}
	score^{spar}\left( v_i \right) =\left\| \mathbf{a}_i-\mathbf{a}_{i}^{spar} \right\| _F
\end{equation}
where $\left\| \cdot \right\| _F$ denotes the Frobenius norm of a vector. 

\subsubsection{Random Walk with Restart}

The Contrastive Learning-based method mainly calculates the agreement between the target node and its surrounding community. Therefore, we adopt the RWR (Random Walk with Restart) to sample the local subgraph with the size of $P$ for each node. The subgraph sampled from target node $v_i$is regarded as the positive pair, which is denoted as $\mathcal{G} _{i}^{dense}$ or $\mathcal{G} _{i}^{spar} $depending on sampled from which view. By contrast, the subgraph sampled from the other node is regarded as the negative pair, the same, it is denoted by $ \tilde{\mathcal{G}}_{i}^{dense} $ or $ \tilde{\mathcal{G}}_{i}^{spar}$. Furthermore, in order to facilitate the following narrative, the similarity information vector of node $v_i$ is defined as $ \mathbf{s}_i \in \mathbb{R} ^P$. Each element of the vector corresponds to the similarity score between the target node and each node in the subgraph respectively.

\begin{algorithm}[tb]
	\caption{Proposed model SCALA}
	\label{alg:algorithm}
	\textbf{Input:} Attributed network $\mathcal{G}=\left(\mathcal{V},\mathcal{E},\mathbf{X}\right)$; Number of training epoch \textit{T}; Batch size \textit{B}; Number of sampling rounds \textit{R}.\\
	\textbf{Output:} Anomaly score mapping function:$score$($\cdot$).
	
	\begin{algorithmic}[1] 
		\STATE Calculate the similarity matrix and leverage the sparisification via Eq.(1) and (2).
		\STATE // \textit{Training phase}.
		\FOR{ \textit{t} $\in \left\{1,2,\cdots,\textit{T}\right\}$ }
		\STATE $\mathcal{B}$ $\gets$ Randomly split $\mathcal{V}$ into batches of size B.
			\FOR{ \textit{b} $\in$ $\mathcal{B}$ }
			\STATE Sample the positive and negative pairs of target nodes in \textit{b} from both \textit{dense-view} and \textit{spar-view}.
			\STATE Calculate the discriminative score in both view via Eq.(4)-(13).
			\STATE Calculate the $\mathcal{L}$ via Eq.(14) and (15).
			\STATE Execute back propagation and update the parameters of contrastive learning model.
			\ENDFOR
		\ENDFOR
		\STATE //  \textit{Inference phase}.
		\FOR{ $v_i \in \mathcal{V}$} 
		\STATE Sample \textit{R} positive and negative pairs of target nodes from both \textit{dense-view} and \textit{spar-view}, respectively.
		\STATE Calculate the anomaly score $score$($v_i$) via Eq.(16)-(19).
		\ENDFOR
	\end{algorithmic}
	
\end{algorithm}

\subsection{Contrastive Learning Module}

In this section, we introduce the contrastive module of SCALA. The contrastive module contains two views. The node-graph contrast method is employed on both views which  aims to calculate the discriminative score. The main difference between the two views is that different readout strategies are adopted. Furthermore, in order to prevent the influence on the discrimination from target node information in the subgraph, the anonymization is conducted on the attribute of the node in the subgraph corresponding to the target node.

\subsubsection{Dense-View Contrastive Learning}
The overall framework in both views is similar in general. First, for the subgraph $\mathcal{G} _{i}^{dense}$ of the target node $v_i$ we apply a GCN mapping the attribute to the low dimensional embedding, which can be formulated as:
\begin{equation}
\label{eq1}
	\mathbf{H}_{i}^{l}=\phi \left( \tilde{\mathbf{D}}_{i}^{-\frac{1}{2}}\tilde{\mathbf{A}}_i\tilde{\mathbf{D}}_{i}^{-\frac{1}{2}}\mathbf{H}_{i}^{\left( l-1 \right)}\mathbf{W}^{\left( l-1 \right)} \right)
\end{equation}
where $\mathbf{H}_{i}^{l}$ and $\mathbf{H}_{i}^{\left( l-1 \right)}$ means the $l^{th}$ and the $\left( l-1 \right) ^{th}$ layer embeddings of subgraph, $\tilde{\mathbf{D}}_i$ is the degree matrix of $\tilde{\mathbf{A}}_i=\mathbf{A}_i+\mathbf{I}_P$,  the $\mathbf{W}^{\left( l-1 \right)}$ is the learnable parameter of the network and the $\phi \left( \cdot \right) $ is the PReLU activation function. Then we employ an MLP that shares the same parameter with the above GCN. This can be formulated as:
\begin{equation}
\label{eq1}
	\mathbf{h}_{i}^{l}=\phi \left( \mathbf{h}_{i}^{\left( l-1 \right)}\mathbf{W}^{\left( l-1 \right)} \right) 
\end{equation}
where $\mathbf{h}_{i}^{l}$ and $\mathbf{h}_{i}^{\left( l-1 \right)}$ is the $l^{th}$ and $\left( l-1 \right) ^{th}$  layer embedding of target node $v_i$, respectively. In order to obtain the latent representation of the subgraph, we leverage an Average Readout function which can fully exploit all the embeddings of the subgraph:
\begin{equation}
\label{eq1}
	\mathbf{e}_{i}^{d}=\sum_{k=1}^P{\frac{\left( \mathbf{H}_i \right) _k}{P}} 
\end{equation}
where the $\mathbf{H}_i\in \mathbb{R} ^{P\times d}$ is the output of the GCN, $\mathbf{e}_{i}^{d}\in \mathbb{R} ^d$ represents the graph-level embedding of $\mathcal{G} _{i}^{dense}$ and $d$ is the dimension of the final embedding. Similarly, the embedding of negative pair subgraph $\tilde{\mathbf{e}}_{i}^{d}$ can obtained by the same procedure. The agreement score between the target node and the subgraph is calculated by a Bilinear Discriminator:
\begin{equation}
\label{eq1}
    \begin{aligned}
    s_{i}^{d}&=Discrimintor\left( \mathbf{h}_i, \mathbf{e}_{i}^{d} \right)\\
    &=\sigma \left( \mathbf{h}_i\mathbf{W}_d\left( \mathbf{e}_{i}^{d} \right) ^T \right)\\
    \tilde{s}_{i}^{d}&=\sigma \left( \mathbf{h}_i\mathbf{W}_d\left( \tilde{\mathbf{e}}_{i}^{d} \right) ^T \right)\\
\end{aligned}
\end{equation}
where $\mathbf{W}_d\in \mathbb{R} ^{d\times d}$ is the parameter of the discriminator and the $\sigma \left( \cdot \right) $ is the sigmoid activation function that maps the score into range $\left[ 0, 1 \right] $. Following the CoLA, the discriminative score of positive pair $s_{i}^{d}$ is supposed to close to $1$, by contrast, the discriminative score of negative pair $\tilde{s}_{i}^{d}$ is supposed to close to $0$. Thus, the BCE (Binary Cross Entropy) loss function is employed to train the module, the loss function of \textit{dense-view} can be :
\begin{equation}
\label{eq1}
	\mathcal{L} ^{dense}=-\frac{1}{2}\left( \log \left( s_{i}^{d} \right) +\log \left( 1-\tilde{s}_{i}^{d} \right) \right)
\end{equation}

\subsubsection{Spar-View Contrastive Learning}

To mitigate the effect of anomalous relations on obtaining graph embeddings, we creatively introduce the \textit{spar-view} to SCALA. As in the \textit{dense-view}, we likewise employ a GCN in order to obtain the embedding $\hat{\mathbf{H}}_{i}^{l}$ of subgraph $\mathcal{G} _{i}^{spar} $. Similarly, to access the embedding $\hat{\mathbf{h}}_{i}^{l}$ of the target node, we employ an MLP with the same learnable weight of the GCN:
\begin{equation}
    \label{eq1}
    \hat{\mathbf{H}}_{i}^{l}=\phi \left( {\tilde{\mathbf{D}}_{i}^{*}}^{-\frac{1}{2}}\tilde{\mathbf{A}}_{i}^{*}{\tilde{\mathbf{D}}_{i}^{*}}^{-\frac{1}{2}}\hat{\mathbf{H}}_{i}^{\left( l-1 \right)}\hat{\mathbf{W}}^{\left( l-1 \right)} \right) 
\end{equation}
\begin{equation}
\label{eq1}
	\hat{\mathbf{h}}_{i}^{l}=\phi \left( \hat{\mathbf{h}}_{i}^{\left( l-1 \right)}\hat{\mathbf{W}}^{\left( l-1 \right)} \right) 
\end{equation}
where  ${\tilde{\mathbf{D}}_{i}^{*}}$ is the degree matrix of $ \tilde{\mathbf{A}}_{i}^{*}=\mathbf{A}^{spar}+\mathbf{I}_P$. To minimize the influence on graph-level embedding by noise information, we no longer use the unselected Average Readout and propose a simple yet effective similarity-based attention mechanism in \textit{spar-view}, which can adaptively select useful embedding of the subgraph. Intuitively, if the node in the subgraph is more similar to the target node, the more weight it should have in the process of pooling and vice versa. In order to allow the module to adaptively utilize the similarity information, an MLP is leveraged to apply a homogeneous dimensional mapping to similarity vector $\mathbf{s}_i $, which can be formulated as:
\begin{equation}
\label{eq1}
	\mathbf{s}_{i}^{attr}=\sigma \left( \mathbf{s}_i\mathbf{W}_s+\mathbf{b} \right)
\end{equation}
where $\mathbf{W}_s\in \mathbb{R} ^{P\times P}$ is the learnable parameter of the MLP. $\mathbf{s}_{i}^{attr}$ is obtained as attention mechanism factor  from the similarity vectors $\mathbf{s}_i $ and applied to the process of obtaining graph embedding:
\begin{equation}
\label{eq1}
	\mathbf{e}_{i}^{s}=\mathbf{s}_{i}^{attr}\hat{\mathbf{H}}_i
\end{equation}

\begin{figure*}[t]
	\centering
	\setlength{\belowcaptionskip}{-0.25cm}
	\begin{subfigure}{0.3\linewidth}
		\centering
		\includegraphics[width=1\linewidth,trim=0 0 0 50,clip]{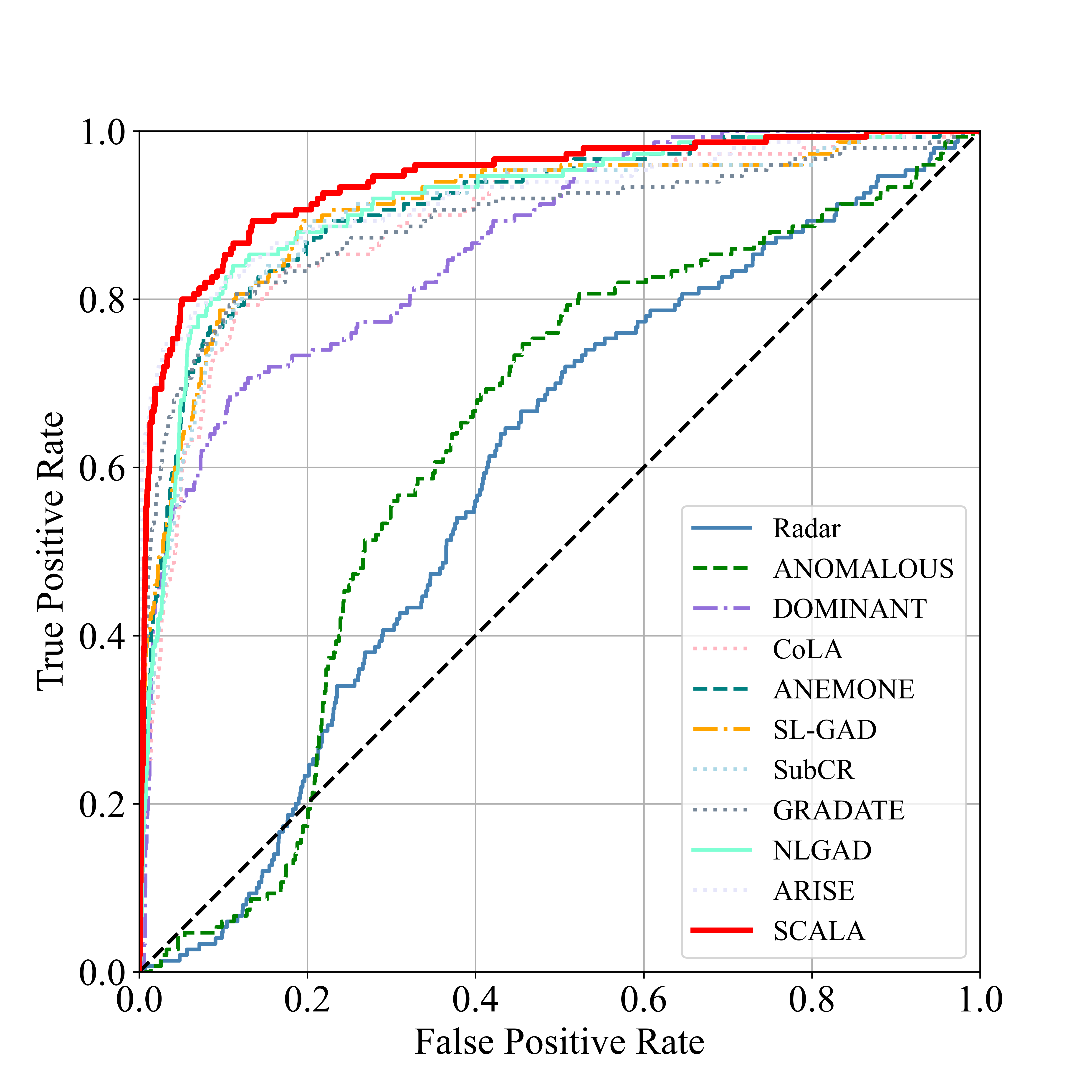}
		\caption{Cora}
	\end{subfigure}
	\centering
	\begin{subfigure}{0.3\linewidth}
		\centering
		\includegraphics[width=1\linewidth,trim=0 0 0 50,clip]{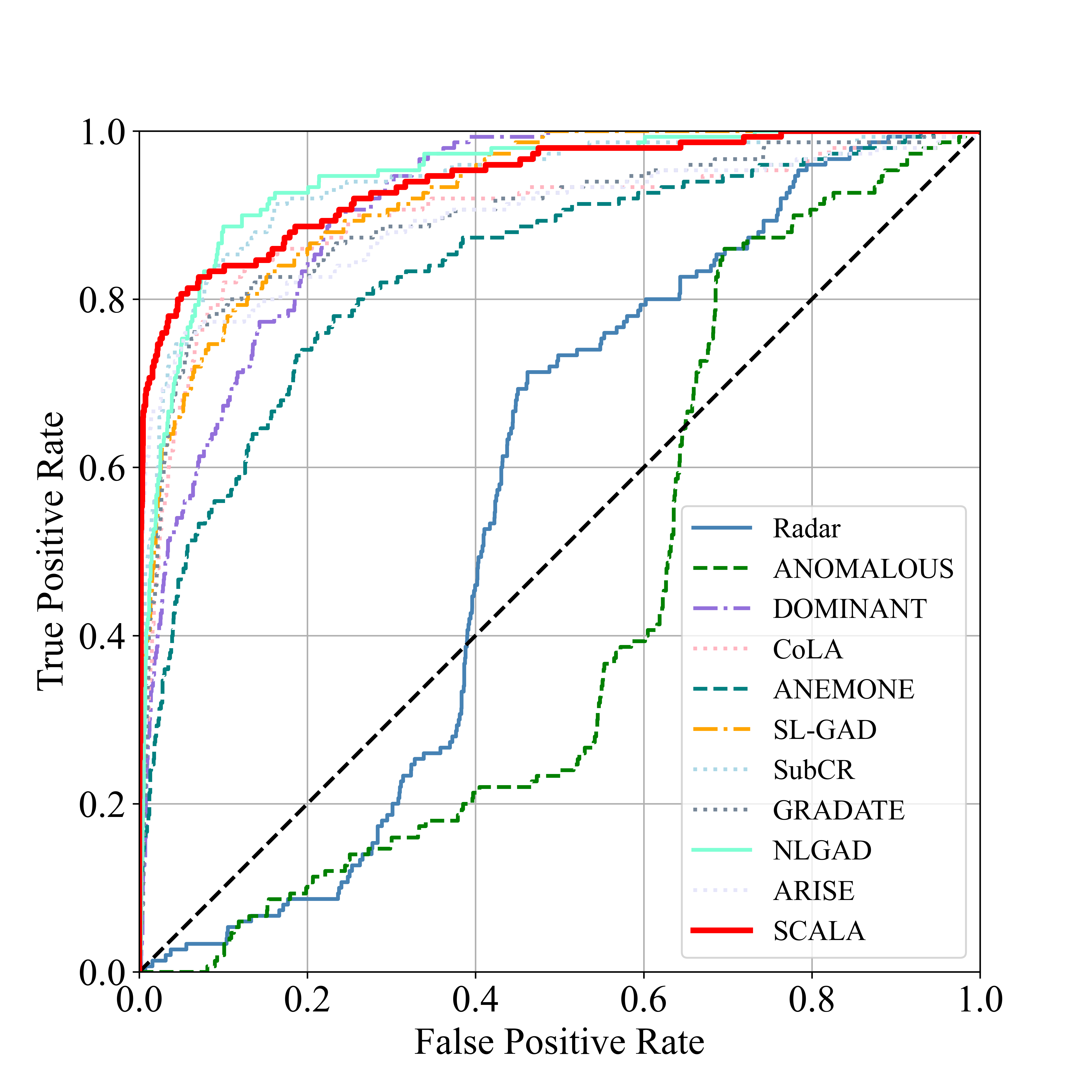}
		\caption{CiteSeer}
	\end{subfigure}
	\centering
	\begin{subfigure}{0.3\linewidth}
		\centering
		\includegraphics[width=1\linewidth,trim=0 0 0 50,clip]{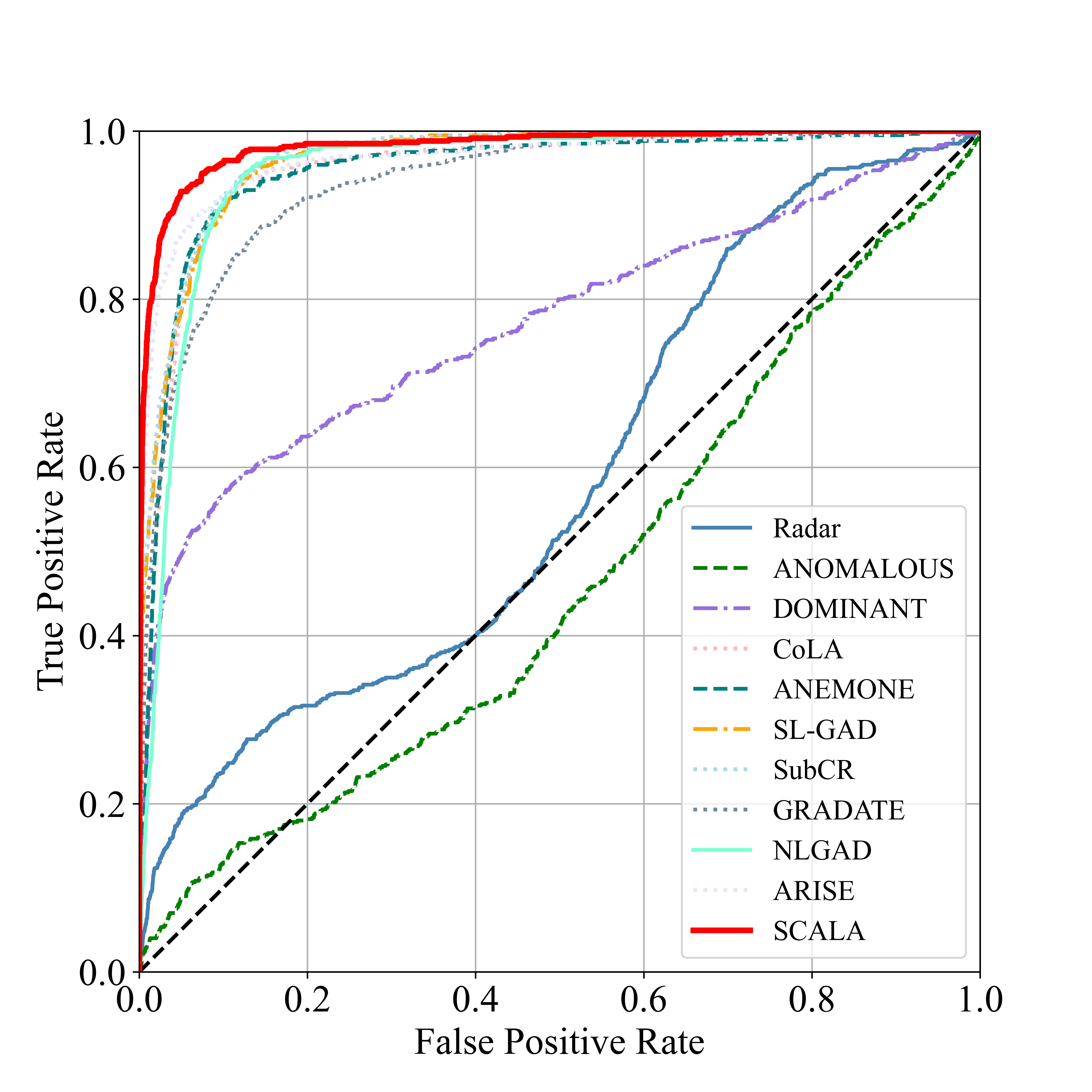}
		\caption{PubMed}
	\end{subfigure}
	\begin{subfigure}{0.3\linewidth}
		\centering
		\includegraphics[width=1\linewidth,trim=0 0 0 50,clip]{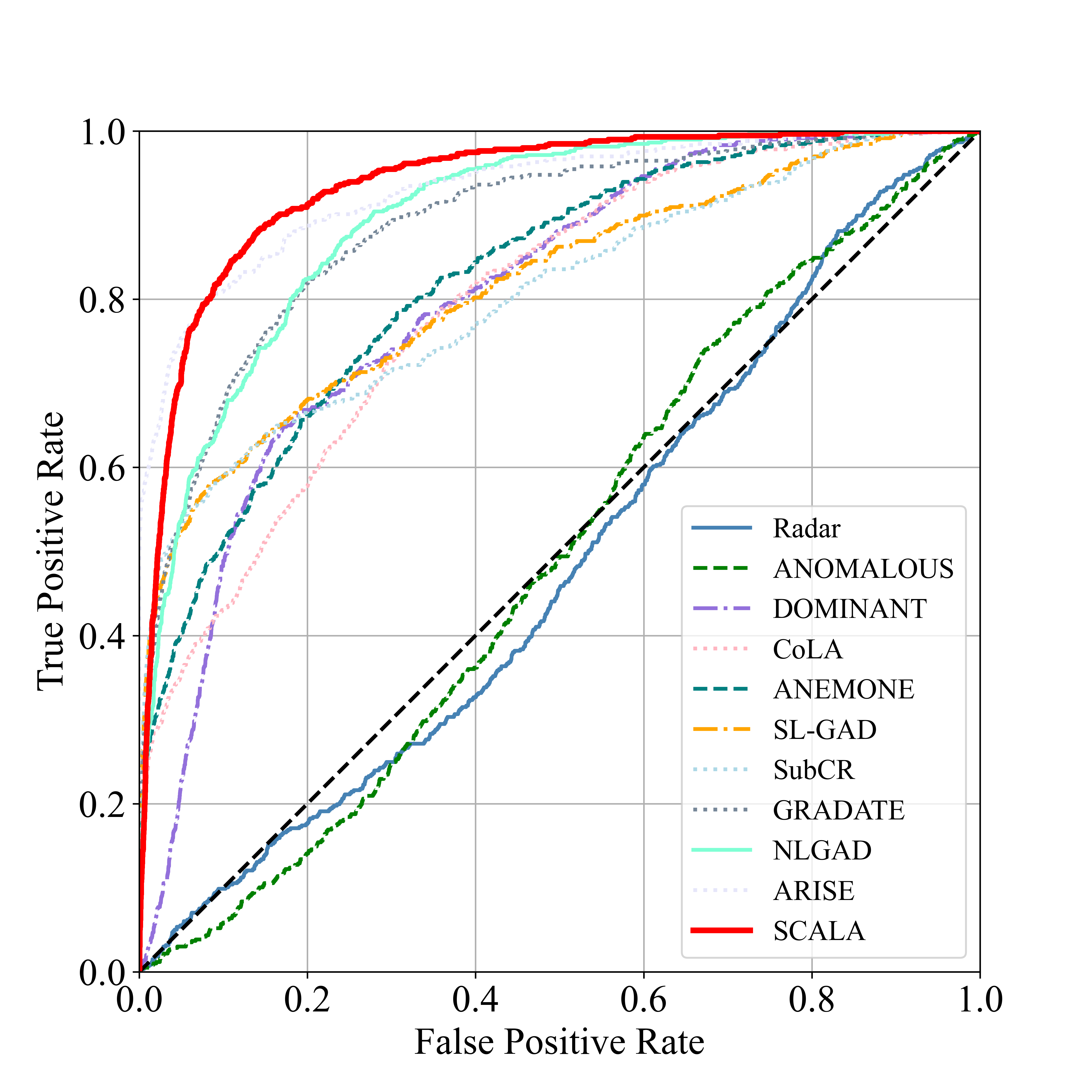}
		\caption{ACM}
	\end{subfigure}\centering
	\begin{subfigure}{0.3\linewidth}
		\centering
		\includegraphics[width=1\linewidth,trim=0 0 0 50,clip]{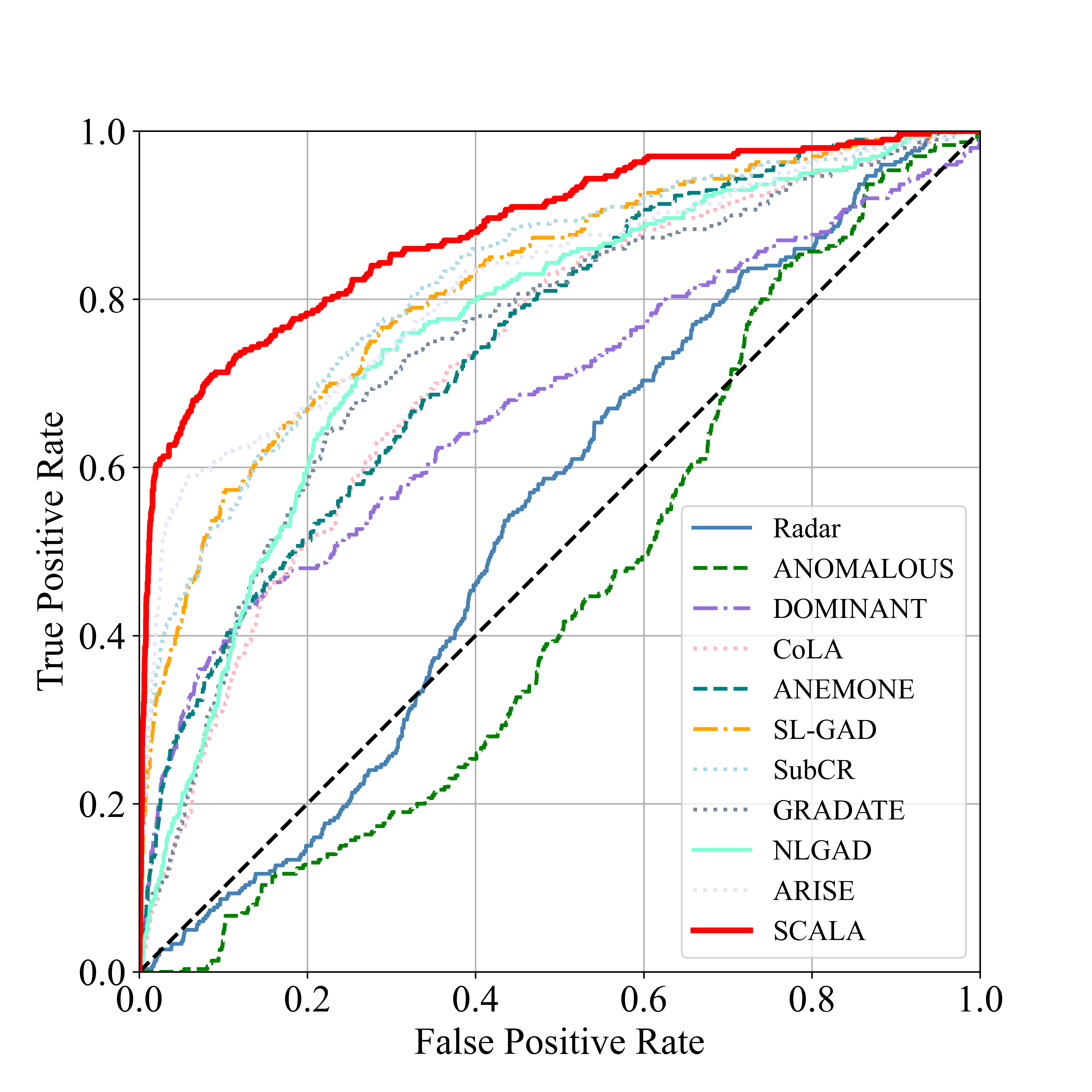}
		\caption{DBLP}
	\end{subfigure}
	\caption{ROC curves on five benchmark datasets.}
	\label{fig: Figure 3}
\end{figure*}

As mentioned before, both views adopt a node-subgraph contrastive method. Therefore, both views will use the same discriminator together. The additional advantage is that the information interaction between the two views is effectively enhanced:
\begin{equation}
\label{eq1}
	\begin{aligned}
	s_{i}^{s}&=\sigma \left( \mathbf{h}_i\mathbf{W}_d\left( \mathbf{e}_{i}^{s} \right) ^T \right)\\
	\tilde{s}_{i}^{s}&=\sigma \left( \mathbf{h}_i\mathbf{W}_d\left( \tilde{\mathbf{e}}_{i}^{s} \right) ^T \right)\\
\end{aligned}
\end{equation}
As with \textit{dense-view}, the BCE is employed as the loss function:
\begin{equation}
\label{eq1}
	\mathcal{L} ^{spar}=-\frac{1}{2}\left( \log \left( s_{i}^{s} \right) +\log \left( 1-\tilde{s}_{i}^{s} \right) \right)
\end{equation}
Ultimately, to train the contrastive learning module for both views simultaneously, the final loss function is defined as:
\begin{equation}
\label{eq1}
	\mathcal{L} =\left( 1-\gamma \right) \mathcal{L} ^{dense}+\gamma \mathcal{L} ^{spar}
\end{equation}
where $\gamma $ is the factor to balance the importance between two views.

\begin{table}[t]
	\centering
	\setlength{\belowcaptionskip}{-0.4cm}
	\resizebox{1\linewidth}{!}{
		\begin{tabular}{c|cccc}
			\toprule 
			Dataset  & Nodes & Edges & Attributes & Anomalies \\
			\midrule
			Cora       & 2708      & 5429    & 1433   &  150  \\
			Citeseer   & 3327      & 4732    & 3703   &  150  \\
			PubMed     & 19717     & 44338   & 500    &  600  \\
			ACM        & 16484     & 71980   & 8337   &  600  \\
			DBLP       & 5484      & 8117    & 6775   &  300  \\
			\bottomrule
		\end{tabular}
	}
	\caption{The statistics of the datasets.}
	\label{tab: Table 1}
\end{table}

\subsection{Anomaly Score Inference}

After training the contrastive learning network, the goal now is to compute the anomaly score for each node. Compared with the normal nodes, it's harder for the model to distinguish the discriminative scores between positive pair and negative pair of the anomaly node. In accordance with CoLA, the score measured by two contrastive modules can be formulated as:
\begin{equation}
\label{eq1}
	score^{con-d}\left( v_i \right) =\tilde{s}_{i}^{d}-s_{i}^{d}
\end{equation}
\begin{equation}
\label{eq1}
	score^{con-s}\left( v_i \right) =\tilde{s}_{i}^{s}-s_{i}^{s}
\end{equation}
The anomaly scores of the two views are then unified by the coefficient $\gamma$:
\begin{equation}
\label{eq1}
	score^{con}\left( v_i \right) =\left( 1-\gamma \right) score^{con-d}+\gamma score^{con-s}
\end{equation}
Notice that it is difficult to adequately capture the context with respect to the target node with only one conduction of sampling. Therefore, multiple samples were taken during the anomaly score inference process and the average of these scores was used as the final score for the whole contrastive module:

\begin{table*}[t]
	\centering
	\footnotesize
	\setlength{\belowcaptionskip}{-0.4cm}
	\setlength{\tabcolsep}{12pt}
	\begin{tabular}{c|ccccc}
		\toprule
		Method   & Cora &  CiteSeer & PubMed  & ACM & DBLP \\
		\midrule
		Radar$^{[2017]}$          & 0.5906      & 0.5580    & 0.5813   & 0.4848   &  0.5411   \\
		ANOMALOUS$^{[2018]}$      & 0.6279      & 0.4336    & 0.4624   & 0.4967   & 0.4508    \\
		DOMINANT$^{[2019]}$       & 0.8639      & 0.9112    & 0.7709   & 0.8009   & 0.6780    \\
		CoLA$^{[2021]}$           & 0.8910      & 0.8982    & 0.9532   & 0.7957   & 0.7291    \\
		ANEMONE$^{[2021]}$        & 0.9096      & 0.8356    & 0.9527   & 0.8226   & 0.7474    \\
		SL-GAD$^{[2021]}$         & 0.9080      & 0.9243    & 0.9662   & 0.8156   & 0.8170    \\
		Sub-CR$^{[2022]} $        & 0.9018      & 0.9385    & \underline{0.9687}   & 0.8051   & 0.8224  \\
		GRADATE$^{[2023]}$        & 0.9053      & 0.8978    & 0.9547   & 0.8881   & 0.7482  \\
		NLGAD$^{[2023]}$          & 0.9173      & \underline{0.9446}   & 0.9538   & 0.8977  & 0.7762  \\
		ARISE$^{[2023]}$          & \underline{0.9226}     & 0.8966   & 0.9664   & \underline{0.9217}   & \underline{0.8278}  \\
		\midrule
		\textbf{SCALA}           & \textbf{0.9492}  & \textbf{0.9710}  & \textbf{0.9847}  & \textbf{0.9442}  & \textbf{0.8849} \\
		\bottomrule
	\end{tabular}
	\caption{Performance comparison for AUC. The bold and underlined values indicates the best and the under-best results, respectively}
	\label{tab: Table 2}
\end{table*}

\begin{equation}
\label{eq1}
    score^{con}\left( v_i \right) =\frac{1}{R}\sum_{r=1}^R{score_{r}^{con}\left( v_i \right)}
\end{equation}
where $R$ is the total number of sampling rounds. Ultimately, the scores from graph sparsification module and contrastive learning module are aggregated as the final score of SCALA:
\begin{equation}
\label{eq1}
	score\left( v_i \right) =\left( 1-\lambda \right) score^{con}\left( v_i \right) +\lambda score^{spar}\left( v_i \right)
\end{equation}
where $\lambda $ is the trade-off parameter of valuing the percentage of each of the two component scores.

\section{Experiment}
In this section, we conduct extensive experiments on five network benchmark datasets that are widely used in anomaly detection on attributed networks to verify the performance of SCALA.

\subsection{Experiment Setting}

\subsubsection{Datasets}

To comprehensively evaluate the proposed model, we choose five benchmark datasets including Cora, Citeseer, PubMed \cite{cora}, ACM, and DBLP \cite{acmdblp}. Due to the shortage of ground truth anomalies in these datasets, we refer to the perturbation scheme following DOMINANT to inject structural anomalies and attribute anomalies for each dataset \cite{dominant,strutural,attribute}. The statistics of the datasets are shown in Table \ref{tab: Table 1}.

\subsubsection{Baselines and Evaluation Metrics}
In this subsection, we compare SCALA with ten well-known baseline methods. The first two models (Radar and ANOMALOUS \cite{radar,anomalous}) are shallow algorithms, and the rest are based on graph neural networks, including a Auto-Encoder based method (DOMINANT \cite{dominant}), five CL-based methods (CoLA, ANEMONE, GRADATE, NLGAD, and ARISE \cite{cola,anemone,gradate,nlgad,arise}), and two hybrid methods that combine Auto-Encoder and CL based learning (SL-GAD and Sub-CR \cite{slgad,subcr}). ROC-AUC is utilized to measure the performance which is a widely-used anomaly detection metric.
%

\subsubsection{Implementation Details}

In our experiments, the size of subgraph $P$ in the network is set to 4 for all datasets by considering both efficiency and performance. The one-layer GCN is employed as the encoder on both views and the embedding dimension is set to 64. The model is optimized with the Adam optimizer during training. The batch size is set to 300 for all datasets. The learning rate is set to 0.001 for Cora, Citeseer, and PubMed, set to 0.0005 for ACM, and set to 0.003 for DBLP. We train the model 100 epochs for Cora, Citeseer and PubMed, and 400 epochs for ACM and DBLP. In the inference phase, we set the number of rounds $R$ to 256 to get the final anomaly score for each node.

\subsection{Result Analysis}

In this subsection, we compare SCALA with ten baseline methods. Figure \ref{fig: Figure 3} demonstrates the ROC curves for 11 models in five datasets. The AUC of SCALA and baselines are summarized in Table \ref{tab: Table 2}. For the results, we have the following conclusions.

We can intuitively find that our proposed model is superior to its competitors on these five datasets. To be specific, SCALA achieves notable AUC gains of 2.66\%, 2.64\%, 1.60\%, 2.25\%, and 5.71\% on Cora, Citeseer, PubMed, ACM, and DBLP, respectively. As shown in Figure \ref{fig: Figure 3}, the AUC of SCALA is significantly larger than others.
 
Shallow methods perform worse than other models due to their limited capability to discriminate anomalies from graphs with high-dimensional features and complex structures. The DOMINANT also does not show competitive performance because it aims to recover attribute and structural information rather than directly capture the anomaly information.
Compared with previous CL-based methods, CoLA, ANEMONE, GRADATE, NLGAD, and ARISE, SCALA has significant improvements on five datasets. It indicates that the sparsification and the attention mechanism can effectively improve the quality of graph-level embedding.
Compared with two hybrid methods that combine reconstruction-based and contrastive-based modules (Sub-CR and SL-GAD), SCALA performs better in anomaly detection. This shows that the support of anomaly measurement by sparsification is stronger than attribute reconstruction based on neighbors.

\subsection{Ablation Study}

In order to confirm the effectiveness of each component in SCALA, a series of ablation studies are conducted. The results are presented in Table \ref{tab: Tabel 3}. The w/o Spar and w/o Con denote the model without sparsification and contrastive module, respectively. The w/o \textit{spar-view} denotes the model without \textit{spar-view}.  The w/o weight represents removing the trade-off parameter $\lambda$. 

It can be seen that the full SCALA achieves the best performance, which validates the effectiveness of combining graph sparsification and contrastive learning in a joint learning manner for graph anomaly detection. Especially, SCALA outperforms w/o Spar by 4.80\%, 4.65\%, 3.68\%, 7.75\%, and 12.95\% on five datasets, respectively, which indicates the graph sparsification module provides strong support to the model. Considering the result of w/o Con, it will be found that the sparsification module itself can play a good role in detection. The result of w/o \textit{spar-view} shows that high-quality embedding will enhance the results of anomaly detection. Due to the different nature of the dataset, the degree of enhancement is also different. At last, the result of w/o weight shows that the scores of the two sections do not provide the same level of support and need to be adjusted according to the actual situation


\begin{table}[t]
    \centering
    \setlength{\belowcaptionskip}{-0.5cm}
    \resizebox{1\linewidth}{!}{
    \begin{tabular}{c|ccccc}
        \toprule
                 & Cora &  Citeseer & PubMed  & ACM & DBLP \\
        \midrule
        \textbf{SCALA}          & \textbf{0.9492}     & \textbf{0.9710}     & \textbf{0.9847}     & \textbf{0.9442}     & \textbf{0.8849 }  \\
        w/o Spar       & 0.9012      & 0.9245   &  0.9479  & 0.8667  &  0.7554    \\  
        w/o Con        & 0.9117      & 0.9504   &  0.9438  & 0.9112  &  0.8596    \\
        w/o \textit{spar-view}  & 0.9397      & 0.9679   &  0.9832  & 0.9421  &  0.8832    \\
        w/o weight     & 0.9357      & 0.9677   &  0.9781  & 0.9253  &  0.8729    \\
        \bottomrule
    \end{tabular}
}
    \caption{The AUC values of ablation study.}
    \label{tab: Tabel 3}
\end{table}

\begin{figure*}[t]
	\centering
	\setlength{\belowcaptionskip}{-0.2cm}
	\begin{subfigure}{0.19\linewidth}
		\centering
		\includegraphics[width=1\linewidth,trim=6 50 10 50,clip]{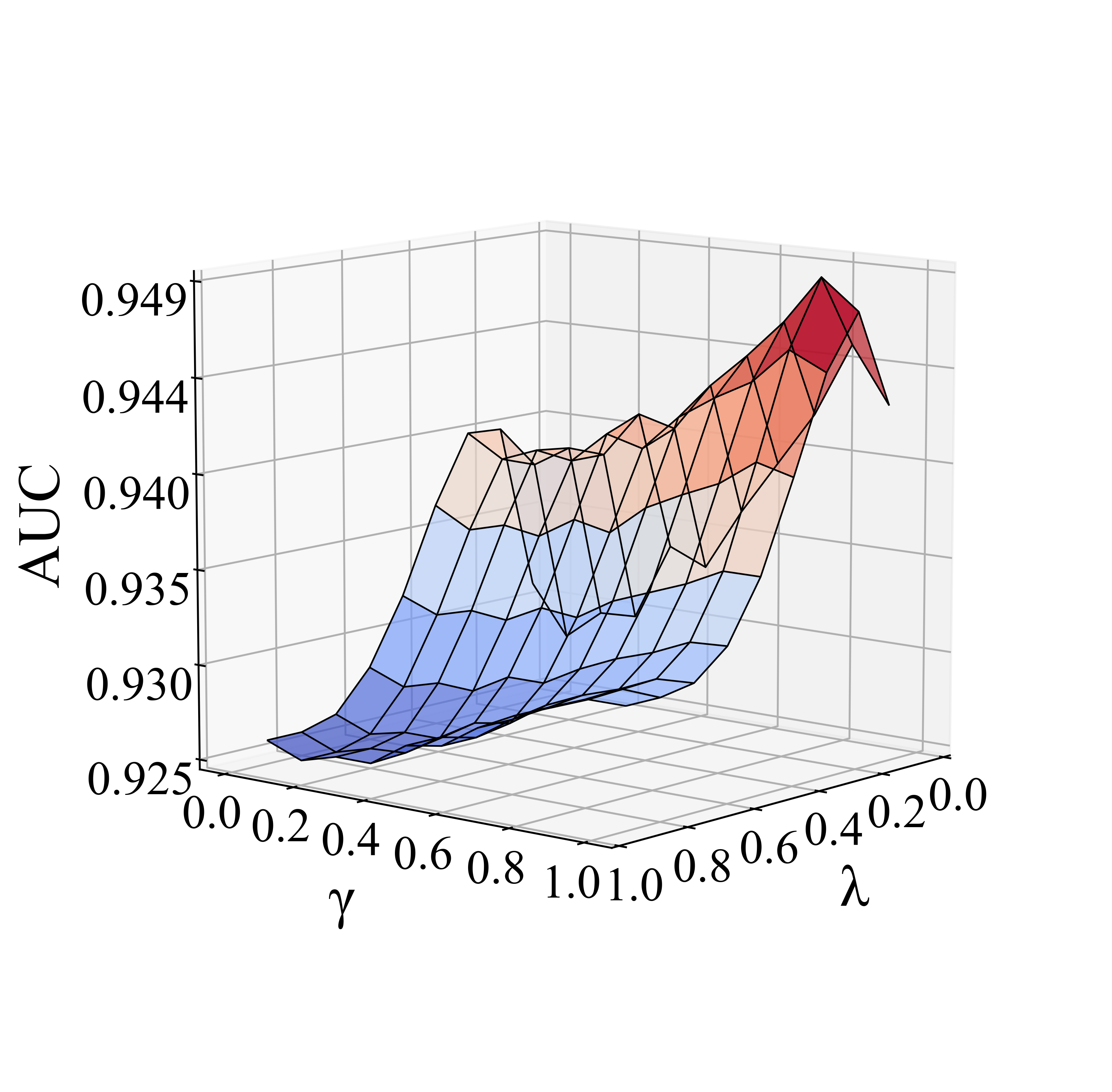}
		\caption{Cora}
		\label{fig:subfig_a}
	\end{subfigure}
	\centering
	\begin{subfigure}{0.19\linewidth}
		\centering
		\includegraphics[width=1\linewidth,trim=6 50 10 50,clip]{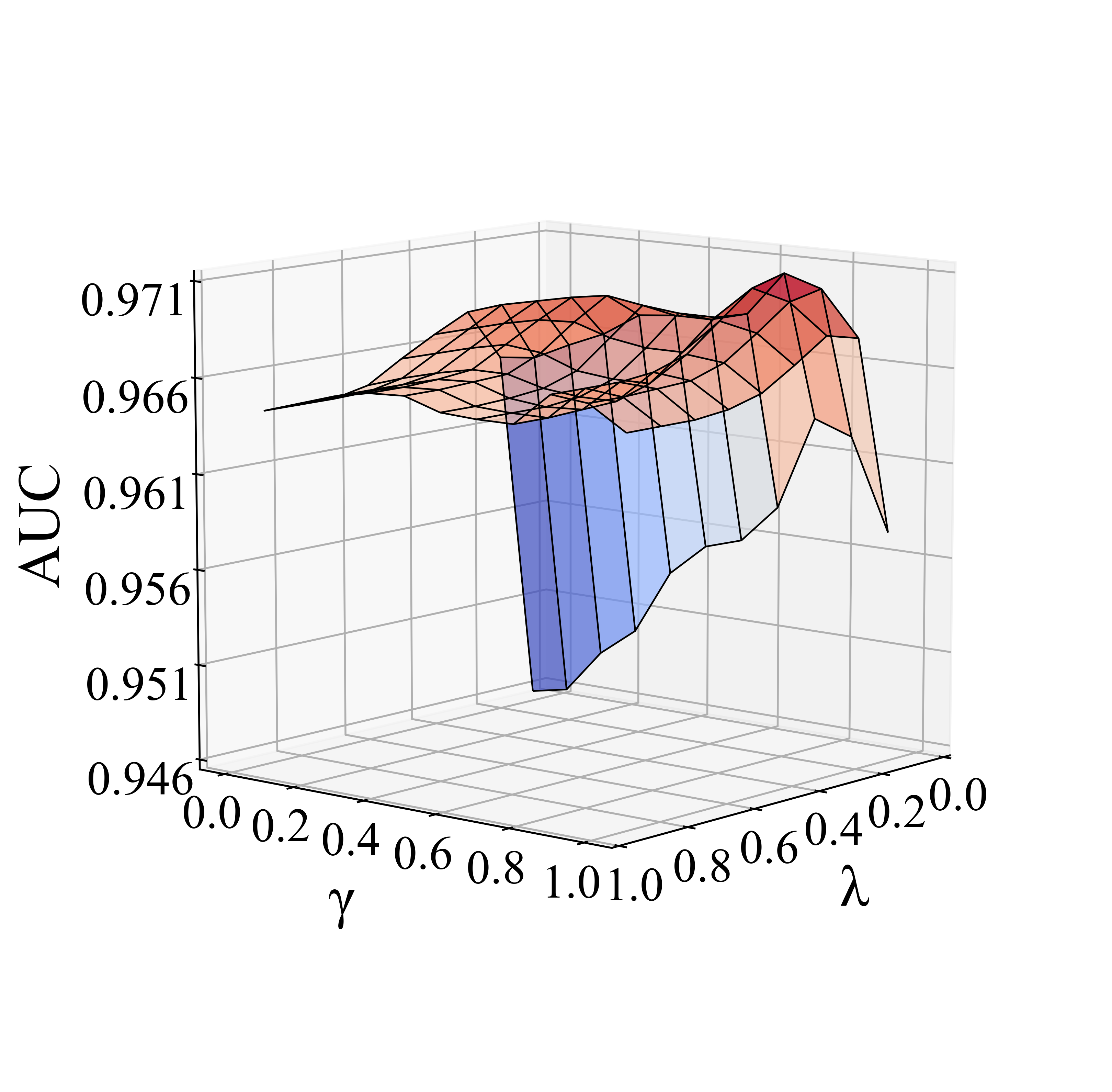}
		\caption{CiteSeer}
		\label{fig:subfig_b}
	\end{subfigure}
	\centering
	\begin{subfigure}{0.19\linewidth}
		\centering
		\includegraphics[width=1\linewidth,trim=6 50 10 50,clip]{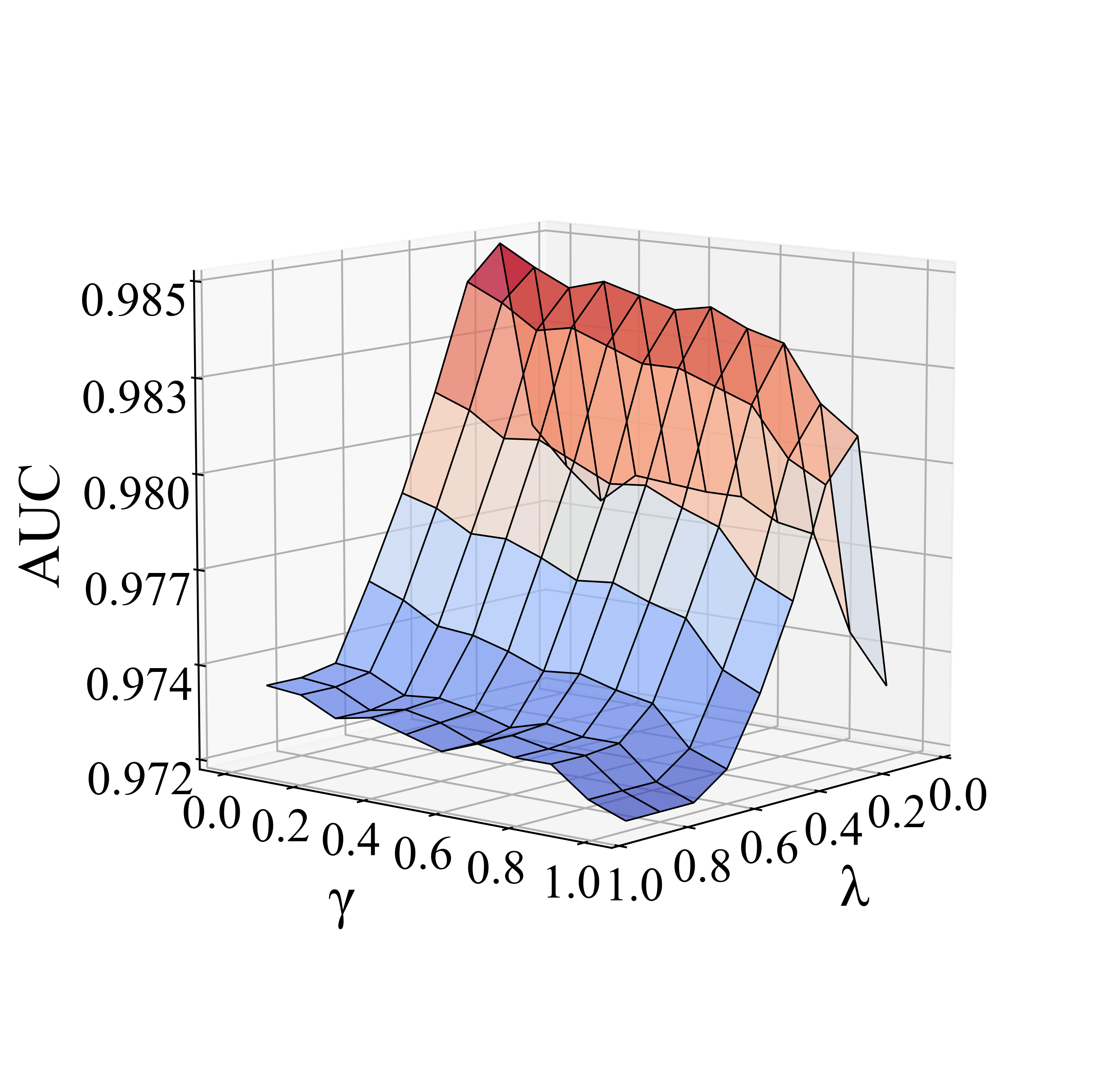}
		\caption{PubMed}
		\label{fig:subfig_c}
	\end{subfigure}
	\centering
	\begin{subfigure}{0.19\linewidth}
		\centering
		\includegraphics[width=1\linewidth,trim=6 50 10 50,clip]{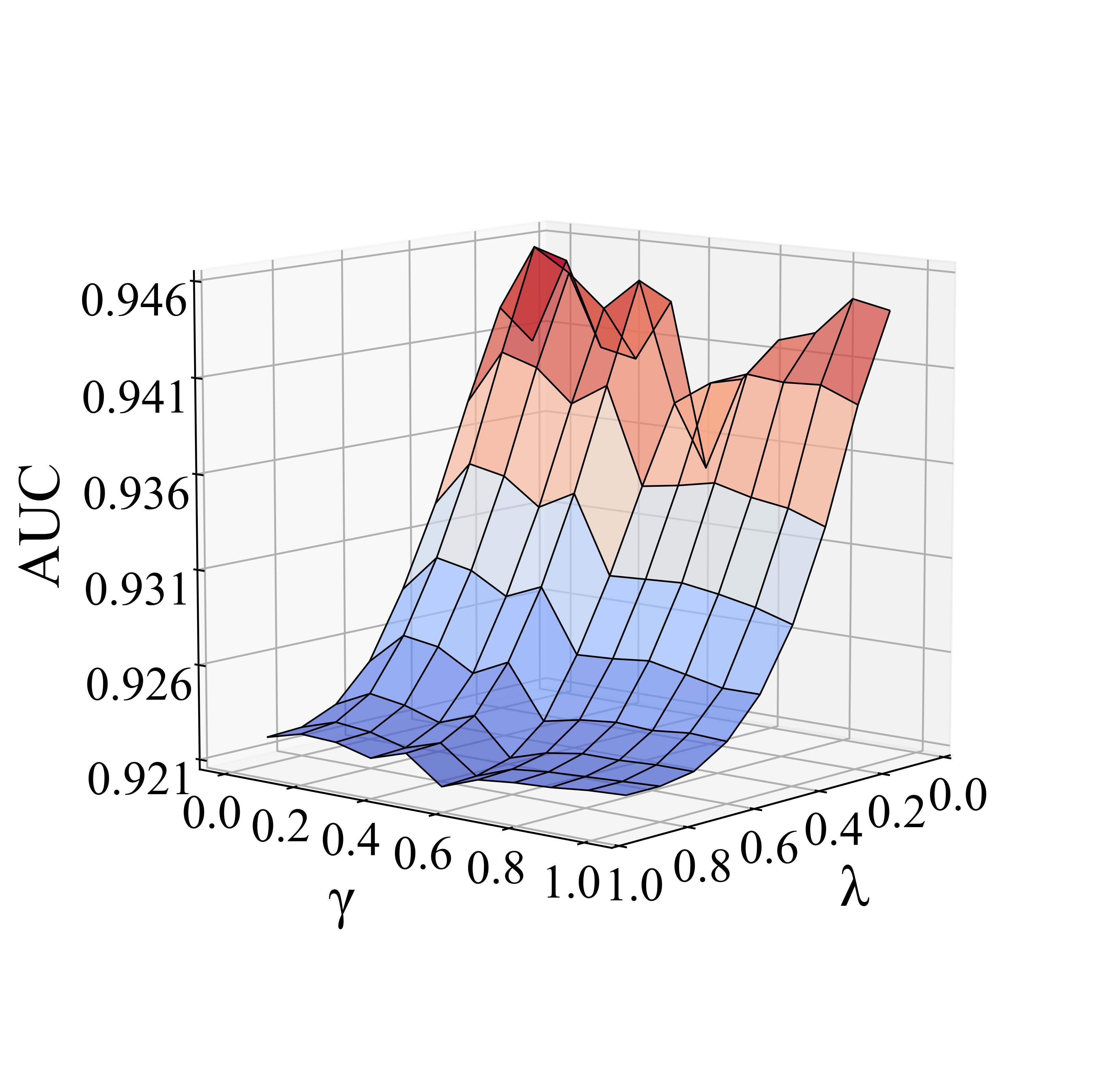}
		\caption{ACM}
		\label{fig:subfig_d}
	\end{subfigure}
	\centering
	\begin{subfigure}{0.19\linewidth}
		\centering
		\includegraphics[width=1\linewidth,trim=6 50 10 50,clip]{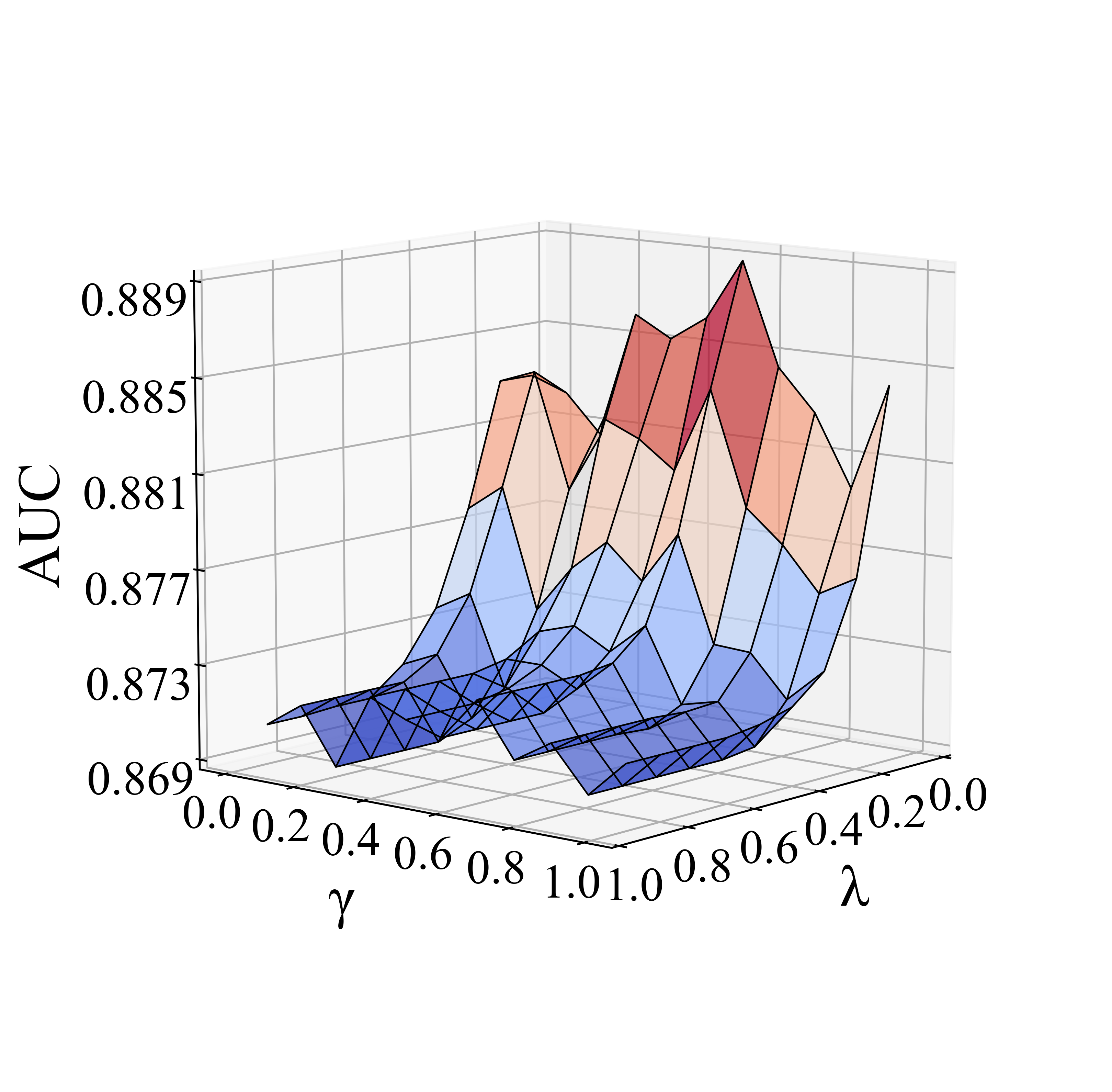}
		\caption{DBLP}
		\label{fig:subfig_e}
	\end{subfigure}
	\centering
	\caption{Performance with different parameters.}
	\label{fig: Figure 4}
\end{figure*}

\subsection{Sensitivity Analysis}

We finally conduct the model sensitivity analysis on critical hyper-parameters in SCALA, which are the balance factor $\gamma$  between \textit{dense-view} and \textit{spar-view}, the coefficient $\lambda$ of anomaly score from two modules, and the threshold $\varepsilon$ of graph sparsification.

Figure \ref{fig: Figure 4} shows the effect of different $\gamma$ and $\lambda$ values on the model performance. We perform a grid search with stride 0.1 for $\gamma$ and $\lambda$. In practice, we set $\gamma$ to 0.9, 0.8, 0.1, 0.1, 0.6 on Cora, Citeseer, PubMed, ACM, and DBLP. Meanwhile, we set $\lambda$ to 0.1 for DBLP, and the remaining four datasets are all set to 0.2.

From the results, we can gain two inspirations. On one hand, the values of $\lambda$ on the five datasets indicate that different datasets focus on different views based on their characteristics. On the other hand, graph sparsification can provide a favorable supplement to anomaly detection based on contrastive learning in anomaly inference.

\begin{figure}[h]
	\centering
	\setlength{\abovecaptionskip}{-0.1cm}
	\includegraphics[width=0.7\linewidth]{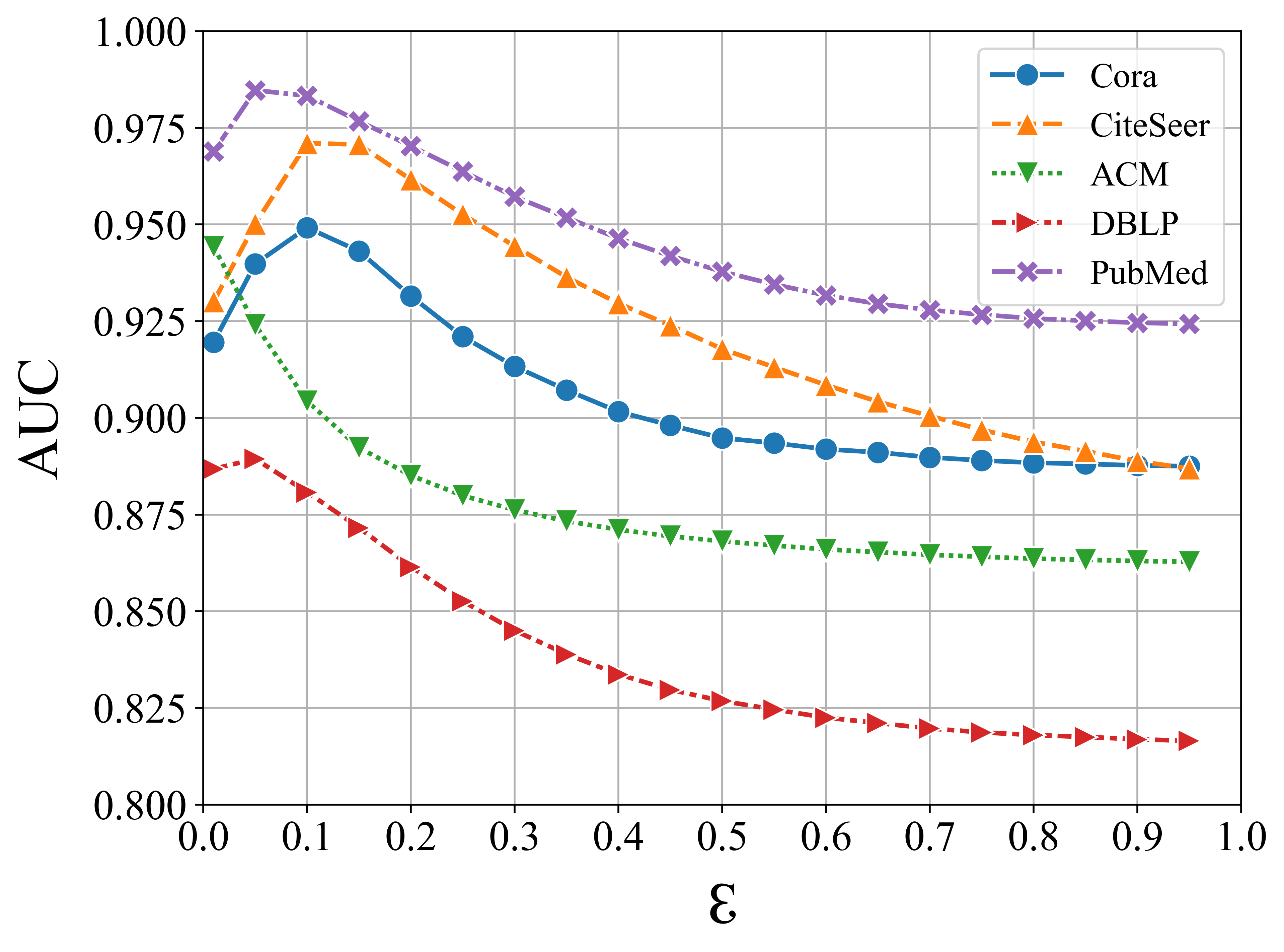}
	\caption{Performance with different $\varepsilon$.}
	\setlength{\belowcaptionskip}{-0.6cm}
	\label{fig: Figure 5}
\end{figure}

Figure \ref{fig: Figure 5} illustrates the performance variation of SCALA when the threshold $\varepsilon$ varies from 0.0 to 1.0. It was observed that the reasonable threshold range is between 0 and 0.1, which means that appropriate graph sparsification based on similarity can effectively filter out some abnormal signals. In practice, the $\varepsilon$ is set to 0.1 for Cora and Citeseer, 0.05 for PubMed and DBLP, and 0.01 for ACM. 

\section{Conclusion}

In this paper, we propose a novel sparsification-based contrastive learning model for anomaly detection on attributed networks. It creatively introduces the sparsification into the procedure, which is not only used as a view augmentation method but also provides a new way of measuring the abnormal. On the new view, we crafted a attention-based readout approach to enhance the quality of graph-level embedding. Extensive experiments have been conducted, demonstrating that the proposed method outperforms its competitors, which convincingly validates the effectiveness of SCALA.

\bibliographystyle{named}
\bibliography{reference}

\end{document}